\documentclass[prd,aps,twocolumn,nofootinbib,showpacs,floats,floatfix]{revtex4}
\usepackage{epsfig}
\usepackage{dcolumn}
\usepackage{bm}
\usepackage{color}
\usepackage{comment}
\usepackage{amsmath,amssymb}
\usepackage[utf8]{inputenc}
\usepackage{soul}

\DeclareGraphicsExtensions{.eps,.ps,.eps.gz,.ps.gz,.eps.Z,{}}

\newcommand{\bv}{{\bf v}}
\newcommand{\bn}{{\bf n}}

\newcommand{\be}{\begin{equation}}
\newcommand{\ee}{\end{equation}}
\newcommand{\bea}{\begin{eqnarray}}
\newcommand{\eea}{\end{eqnarray}}
\newcommand{\bel}{\begin{multline}}
\newcommand{\eel}{\end{multline}}
\newcommand{\al}{\alpha}

\newcommand{\Mc}{\mathcal{M}_c(z)}
\newcommand{\ds}{\displaystyle}

\begin{document}

\title{The effect of matter structure on the gravitational waveform}

\author{Camille Bonvin$^{1}$, Chiara Caprini$^{2,3}$, Riccardo Sturani$^{4}$ and Nicola Tamanini$^{2}$
\\  }
\affiliation{${}^{1}$ D\'epartement de Physique Th\'eorique and Center for Astroparticle Physics (CAP), University of Geneva,
24 quai Ernest Ansermet, CH-1211 Geneva, Switzerland\\
${}^{2}$ Institut de Physique Th\'eorique, CEA-Saclay, CNRS UMR 3681, Universit\'e Paris-Saclay, F-91191 Gif-sur-Yvette, France \\
${}^{3}$ Laboratoire Astroparticule et Cosmologie, CNRS UMR 7164, Universit\'e Paris-Diderot, 10 rue Alice Domon et L\'eonie Duquet, 75013 Paris, France\\
${}^{4}$ Instituto de F\'\i sica Te\'orica UNESP / International Center of Theoretical Physics - South Amertican Institute for Fundamental Research, 01140-070 S\~ao Paulo, Brazil}

\date{\today}

\begin{abstract}

Third generation ground-based interferometers as well as the planned space-based interferometer LISA
are expected to detect a plethora of gravitational wave signals from coalescing binaries at cosmological distance. The emitted gravitational waves propagate in the expanding universe through the inhomogeneous distribution of matter. Here we show that the acceleration of the universe and the peculiar acceleration of the binary with respect to the observer distort the gravitational chirp signal from the simplest General Relativity prediction beyond a mere time independent rescaling of the chirp mass, affecting intrinsic parameter estimations for the binaries visible by LISA. 
We find that the effect due to the peculiar acceleration can be much larger than the one due to the universe acceleration.
Moreover, peculiar accelerations can introduce a bias in the estimation of parameters such as the time of coalescence and the individual masses of the binary.
An error in the estimation of the time of coalescence made by LISA will have an impact on the prediction of the time at which the signal will be visible by ground based 
interferometers, for signals spanning both frequency bands.
\end{abstract}

\maketitle
 
\section{Introduction}
\label{sec:intro}

Gravitational Wave (GW) astronomy has recently started \cite{TheLIGOScientific:2016pea,Abbott:2016blz}, showing that signals from coalescing binaries at cosmological distance (with redshift $z\sim 0.1$) are already a reality and in the near future ($\sim$ few years) dozens of similar signals are expected.

The observational quest for GWs is now lead by earth-based interferometers
\cite{TheLIGOScientific:2014jea,TheLIGOScientific:2016agk}, but in the
future the space-based interferometric
detector LISA is expected to widen the range of detectable
sources up to redshift $z\sim 15$ \cite{elisaweb,Klein:2015hvg}.

GWs from coalescing binaries provide a direct measurement of the luminosity
distance of the source to the observer. However, to first approximation (as long as the variation of the cosmological expansion
can be neglected during the duration of the signal as we will see),
GW observations do not provide
information about the redshift of the source. This happens because the redshift does change the waveform, but
in a way that can be exactly compensated by a shift of the masses from their source to detector values and by replacing the comoving distance with the luminosity distance. It is therefore usually assumed that
the redshift of the host galaxy is needed to infer the redshift of the GW event.
Ref.~\cite{Schutz:1986gp} was the first to show that cosmological parameters like the Hubble constant can be measured with few percent precision with $O(10)$ GW detections, by combining the measurements of the luminosity distances and sky localisations of various GW events with the redshift information taken from galaxy catalogs. Since then, the problem has been widely studied both for advanced earth-based interferometers, e.g.~\cite{Sathyaprakash:2009xt,DelPozzo:2011yh,Taylor:2011fs,Taylor:2012db} and for LISA, e.g.~\cite{VanDenBroeck:2010fp,Arun:2007hu,Arun:2008xf,Tamanini:2016zlh}. 

On the other hand, Ref.~\cite{Seto:2001qf, Nishizawa} showed that the GW observation alone
does in principle allow to measure the real masses \emph{and} the redshift. 
The expansion of the universe during the time of observation of the GW event 
can actually imprint into the waveform phase, to which the 
interferometer output is particularly sensitive, an effect with frequency dependence $f^{-8/3}$ with respect to the leading behavior.
The investigation of the detectability of such effect, hence the possibility of
measuring both the luminosity distance and the redshift from gravitational
wave observations alone has been considered in \cite{Seto:2001qf, Nishizawa}. 

Here we re-analyse the issue, taking into account also the redshift perturbations due to the inhomogeneous matter distribution along the propagation of the GWs from the source to the detector. We show that the peculiar acceleration of the binary (i.e.~the time variation of the peculiar velocity) with respect to the cosmological flow can drown the effect of the expansion of the universe: therefore,
the imprint of the background expansion on the phase of the GW signal cannot be
used in general to infer the redshift of the GW source. Moreover, the peculiar acceleration pollutes the phasing signal introducing a bias in the measured parameters, like the binary constituent masses and the time of coalescence. This can be particularly important for those binaries that are visible first by LISA and afterwards by terrestrial interferometers \cite{Sesana:2016ljz,Barausse:2016eii,Vitale:2016rfr}, for which a precise determination of the arrival time of the signal in the LIGO/Virgo band is needed. 

The main result of this paper is Eq.~\eqref{Psi+}, showing the frequency-dependent modification of the phase of the GW signal due to the {\it time variation} of the redshift perturbations. The paper is structured as follows. In sec.~\ref{se:cosmoredshift} we review the chirp gravitational waveform when the redshift is kept constant and unperturbed. In sec.~\ref{se:astroredshift} we account for time variations of the redshift: first we concentrate on the background effect, due to the variation of the cosmological expansion during the observation time of the binary, and then we present the consequences of the time variation of the redshift perturbations due to the inhomogeneities in the matter distribution at linear order~\footnote{Note that constant perturbations to the redshift do not generate a shift in the phase since they can be reabsorbed in the redshifted chirp mass (see discussion in section~\ref{se:astroredshift}).}. In sec.~\ref{se:waveform} we study the modification of the waveform phasing due to both these effects. In sec.~\ref{se:mismatch} we proceed to a quantitative analysis: first we demonstrate that the most relevant contribution to the waveform comes from the peculiar acceleration of the binary; we then show qualitatively that this is only important for space-based detectors, which are capable to follow the chirp signal for a long enough time at low frequency; at last, we quantify the effect in the output of match-filtering commonly used in GW data analysis, focusing on the case of LISA. We show that the amount of lost detections due to the use of a waveform template without the peculiar acceleration of the binary is negligible. However, the peculiar acceleration introduces a {\it bias} in the determination of the binary parameters such as the time of coalescence and the masses. In sec.~\ref{se:conclusion} we conclude.

Throughout the paper we only consider non-spinning binaries at the lowest Post Newtonian (PN) order (except in the last section, as specified). 
We adopt units such that the speed of light $c=1$. The cosmological metric is $ds^2=-dt^2+a^2\delta_{ij}dx^idx^j$ where $t$ denotes cosmic time, $a(t)$ is the scale factor and $H\equiv\dot a/a$ is the Hubble factor with
$H_0\equiv 100 \, h \,{\rm km/sec/Mpc}$ denoting the Hubble factor today. 

\section{Binary at cosmological distance: unperturbed constant redshift}
\label{se:cosmoredshift}

We consider a binary system with individual masses $m_{1,2}$, total mass
$M\equiv m_1+m_2$, symmetric mass ratio $\eta\equiv m_1 m_2/M^2$ and chirp mass
$M_c\equiv \eta^{3/5}M$. 
The two polarisations of the GW signal emitted by the binary in the transverse-traceless (TT)
gauge are (see e.g.~sec.~4.1 of \cite{michele})
\begin{align}
h_+(t_S)&= \frac{4\left(G M_c\right)^{\frac{5}{3}}}{a(t_S)\,r}
\left(\pi\,f_S\right)^{\frac{2}{3}}\frac{1+\cos^2\imath}{2}\cos[\Phi_S(t_S)]\label{hsource}\\
h_\times(t_S)&= \frac{4\left(G M_c\right)^{\frac{5}{3}}}{a(t_S)\,r}
\left(\pi\,f_S\right)^{\frac{2}{3}}\cos\imath \, \sin[\Phi_S(t_S)] \label{hsourcecross}
\end{align}
where $t_S$ is the proper (retarded) time of the source, $a(t_S)\,r\equiv d_p$ is the proper distance
(the luminosity distance being $d_L=d_p(1+z)$)
and $f_S$ is the frequency of the GW as produced at the source. We have
reported only the leading order in the PN expansion
parameter $x\equiv (\pi G M f_S)^{2/3}$.

The evolution of the GW frequency due to back-reaction
of the emission of GWs at leading-order in $x$
is given by (see e.g.~Eq.~(4.18) of~\cite{michele})
\be
\label{dfs}
\frac{df_S}{dt_S}= C(M_c) f_S^{11/3}
\ee
with $C(M_c)\equiv\frac{96}{5}\pi^{8/3}\left(G M_c\right)^{5/3}$.
The solution to Eq.~\eqref{dfs} is
\be
\label{fvstau}
f_S(\tau_S)=\frac{1}{\pi}\left(\frac{5}{256\tau_S}\right)^{3/8}\left(G M_c \right)^{-5/8}\, ,
\ee
where $\tau_S\equiv t_c-t_S$ is the time to coalescence at the source, when $f_S$ formally diverges.
The phase $\Phi_S$ at the source is then
\be
\Phi_S(t_S)\equiv\ds \Phi_c+2\pi \int_{t_c}^{t_S} dt'_S f_S(t'_S)=-2 \left(\frac{\tau_S}{5GM_c} \right)^{\frac{5}{8}}+\Phi_c\,,
\label{OmS}
\ee
where $\Phi_c$ is the phase at coalescence time $t_c$.

When the GW source is at cosmological distance, the frequency at the observer $f_O$ is related to the one at the source by the
cosmological redshift $z$ via 
\be
\label{f}
f_S=(1+z)f_O\,.
\ee
By definition, the redshift is the quantity that relates the frequency at the source to the one at the observer in any cosmology. 
In a homogeneous and isotropic Friedmann-Robertson-Walker (FRW) universe, the redshift simply becomes 
\be
\label{z}
1+ z \equiv\frac{a_O}{a_S}\,.
\ee 
However, the inhomogeneities and anisotropies of the real universe induce perturbations in the redshift changing the above expression, as we will see in the next section \ref{sec:pert}. 

As for the frequency, the proper time at the source is also related to the proper time at the observer through the redshift
\be
\label{dt}
dt_O=(1+z)dt_S\, .
\ee
Using Eqs.~\eqref{f} and \eqref{dt} we can rewrite Eq.~\eqref{dfs} as
\be
\label{dfO}
(1+z)\frac{d}{dt_O}\Big[(1+z)f_O(t_O) \Big]=C(M_c)f_O(t_O)^{\frac{11}{3}}(1+z)^{\frac{11}{3}}\, .
\ee
Neglecting redshift variations over the emission of the GW (the standard procedure), one obtains
\be
\frac{df_O}{dt_O}= C((1+z)M_c) f_O^{\frac{11}{3}}\,,
\label{eq:dfMchipr}
\ee
showing that the redshift can be absorbed by a re-definition of the chirp mass:
\be
\label{defMc}
\mathcal M_c(z)\equiv (1+z) M_c\,.
\ee
It is then impossible to see the effect of the redshift in the phasing, as it boils down to a shift
in the chirp mass. This degeneracy holds also including higher $x$ corrections to \eqref{dfs} for black holes, i.e. as long as tidal deformation and matter
effects are neglected \cite{Messenger:2011gi}.

The solution of Eq.~\eqref{eq:dfMchipr} is simply
\be
\label{fvotau}
f_O(\tau_O)=\frac{1}{\pi}\left(\frac{5}{256\tau_O}\right)^{3/8}\left(G \mathcal M_c \right)^{-5/8}\,.
\ee
Integrating once we obtain the phase
\be
\label{Phitau}
\Phi_O(\tau_O)=-2 \left(\frac{\tau_O}{5G \mathcal M_c} \right)^{\frac{5}{8}}+\Phi_c\, .
\ee
Eq.~\eqref{Phitau} reflects the fact that the phase is constant during propagation
$\Phi_O(\tau_O(\tau_S))=\Phi_S(\tau_S)$
(this follows directly from the fact that GWs propagate on null geodesics
$k_\mu k^\mu=0$ and that $\partial^\mu \varphi=-k^\mu$). The GW amplitudes
in Eqs.~(\ref{hsource},\,\ref{hsourcecross}) can be rewritten in terms of quantities at the observer:
\bea
h_+(t_O)&=& \frac{4(G \mathcal M_c(z))^{\frac{5}{3}}}{d_L} (\pi f_O)^{\frac{2}{3}}
\frac{1+\cos^2\imath}{2}\, \cos(\Phi_O)\, ,~~~~~ \label{hO}\\
h_\times(t_O)&=& \frac{4(G\mathcal M_c(z))^{\frac{5}{3}}}{d_L} (\pi f_O)^{\frac{2}{3}}
 \cos\imath \,\, \sin(\Phi_O)\, , \label{hOcross}
\eea
where, with respect to Eqs.~(\ref{hsource},\,\ref{hsourcecross}),
the waveform amplitude is proportional to $1/d_L$ instead of $1/d_p$
and has otherwise the same functional form with $M_c$ replaced by
$\mathcal M_c(z)$.

\section{Binary at cosmological distance: varying redshift}
\label{se:astroredshift}

In this section we relax the assumption that the redshift $z$
is constant during the observation of the GW signal.
We can identify two contributions to the variation of $z$:
\begin{enumerate}
\item the background expansion of the universe varies during the time of observation, so that the ratio of the scale factor of the universe at the binary and at the observer vary during the observation, as computed in \cite{Seto:2001qf, Nishizawa}.
\item the perturbations in the redshift due to the inhomogeneities and anisotropies in the distribution of
  matter between the binary and the observer (inside the bracket of eq.~\eqref{dez} below)
  vary during the time of observation. This effect is calculated here for the first time.
  The motivation to look for such an effect is that the perturbations
  (and in particular the peculiar velocity of the binary, as we will see)
  can vary on a shorter time-scale than the expansion of the universe,
  leading to a larger contribution than the one of the background expansion.
\end{enumerate}

We account for linear scalar perturbations in the metric. We work in the longitudinal gauge~\cite{bookRuth}, so that the perturbed metric is
\be
ds^2=-\big(1+2\psi\big)dt^2+a^2\big(1-2\phi\big)\delta_{ij}dx^idx^j\, ,
\ee
where $\psi$ and $\phi$ are the scalar potentials. In this perturbed FRW universe, the redshift can be calculated by solving the null geodesic equation
for the photon momentum $k^\mu$
\be
\frac{dk^\mu}{d\lambda}+\Gamma^\mu_{\al\beta}k^\al k^\beta=0\, ,
\ee
where $\lambda$ is the geodesic affine parameter.
The redshift is then given by the ratio of frequencies, which is equal to the ratio of energies:
\be
1+z=\frac{f_S}{f_O}=\frac{E_S}{E_O}=\frac{\big(k^\mu u_\mu\big)_S}{\big(k^\mu u_\mu\big)_O}\, ,
\ee
where $u_S^\mu$ denotes the four-velocity of the source (respectively the observer)
\be
u^\mu=\frac{1}{a}(1-\psi, \bv)\, ,
\ee
and $\bv$ is the peculiar velocity. At linear order in perturbation theory, one gets (see e.g.~\cite{bookRuth})
\begin{align}
1+z=&\frac{a_O}{a_S}\Bigg[1+\mathbf{v}_S\cdot\mathbf{n}-\mathbf{v}_O\cdot\mathbf{n}\nonumber\\
& +\psi_O-\psi_S   -\int_{t_S}^{t_O}   dt(\dot{\phi}+\dot{\psi}) \Bigg]  \nonumber\\
\equiv&\, (1+\bar z)[1+\delta z] \label{dez}\, ,
\end{align}
where $\bn$ denotes the unit vector pointing from the observer to the binary and a dot is a derivative with respect to physical time $t$. The first term on the right hand side of \eqref{dez} corresponds to the expansion of the background already given in Eq.~\eqref{z}, and we have denoted it with 
\be
1+\bar z\equiv \frac{a_O}{a_S}\,. 
\label{barz}
\ee
The rest of the above expression \eqref{dez} is due to the effect of matter perturbations. The first term is the Doppler redshift due to the difference in peculiar velocity between the source and the observer. The second term is the gravitational redshift generated by the difference in gravitational potential at the position of the source and at the position of the observer. The last term is the so-called integrated Sachs-Wolfe contribution generated by the change in the photons' energy when passing through an evolving gravitational potential. 

These redshift perturbations change the observed frequency of the signal $f_O$ as well as the observed proper time $dt_O$ according to Eqs.~\eqref{f} and~\eqref{dt} (which are also valid in a perturbed universe). However if the perturbations are constant in time, they can be reabsorbed into the redshifted chirp mass~\eqref{defMc} and they do not change the waveform of the signal which is still given by Eqs.~\eqref{Phitau},~\eqref{hO} and~\eqref{hOcross}. 

On the other hand, if the redshift is evolving with time, it cannot be reabsorbed into the chirp mass, as we now show. To solve Eq.~\eqref{dfO} for a non-constant redshift we define a new function
\be
\label{g0}
g(t_O)\equiv\left[1+z(t_O)\right]f(t_O)\, .
\ee
We get
\begin{align}
&\frac{dg_O}{dt_O}= C(M_c)\frac{g_O^{11/3}}{1+z}\, ,\\
&g_O^{-8/3}(t_{O})=\frac{8C(M_c)}{3}\int_{t_{O}}^{t_c}dt'_O\big[1+z(t'_O)\big]^{-1} \,,
\label{dgO}
\end{align}
where $t_c$ is the time of coalescence and we have used that $g_O(t_c)\rightarrow\infty$. We proceed to solve the above equation separating the effect of the expansion of the background (already derived in \cite{Seto:2001qf, Nishizawa}) and the effect of the linear matter perturbations. 

\subsection{Homogeneous universe}

Let us start by re-deriving the effect put forward in \cite{Seto:2001qf, Nishizawa}.
We neglect the perturbations in the redshift but account for the variation of the expansion of the universe, so that from Eq.~\eqref{barz} we have:
\be
\label{dgOi}
g_O^{-8/3}(t_{O})
=\frac{8C(M_c)}{3}\int_{t_{O}}^{t_{c}}dt'_O \frac{a\big(t_S(t'_O)\big)}{a(t'_O)}\, .
\ee
By expanding the scale factor around the value $t_O$,
with $t'_S\equiv t_S(t'_O)$, $t_S\equiv t_S(t_O)$,
\bea
a(t'_O)=a(t_O+t'_O-t_O)\simeq a(t_O)+(t'_O-t_O)\dot{a}(t_O)\, ,~~~&&\\
a(t'_S)=a(t_S+t'_S-t_S)\simeq a(t_S)+(t'_S-t_S)\dot{a}(t_S)\, ,~~~~~&&
\eea
one obtains
\begin{multline}
\label{gO}
g_O^{-8/3}(t_{O})=\frac{a(t_S)}{a(t_O)}\frac{8C(M_c)}{3}\\
\times\int_{t_{O}}^{t_{c}}dt'_O \Big[1+H_S (t'_S-t_S)-H_O(t'_O-t_{O})\Big]\, ,
\end{multline}
where $H\equiv\dot{a}/a$. Neglecting terms of the order $(t'_O-t_O)^2$ in
the integrand we can use $t'_S-t_S=(t'_O-t_O)/(1+\bar z)[1+\mathcal O(H_0(t'_O-t_O))]$
and after integrating the right-hand side we obtain
\be
\label{eq:g0}
g_O^{-8/3}(t_{O})=
\frac{8C(M_c)}{3(1+\bar z)}\Big[\tau_O  -X(z)\tau_O^2 \Big]\, ,
\ee
where (c.f.~definition \eqref{barz})
\be
X(z)\equiv\frac{1}{2}\left(H_0-\frac{H_S}{1+\bar{z}}\right)\,.
\label{eq:Xz}
\ee
Re-expressing the above Eq.~\eqref{eq:g0} in terms of the frequency at the
detector via Eq.~\eqref{g0} one finds (to be compared with Eq.\eqref{fvotau})
\be
\label{fOX}
f_O(\tau_O)=
\frac{1}{\pi}\left(\frac{5}{256\,\tau_O}\right)^{\frac{3}{8}}\left(G \Mc \right)^{-\frac{5}{8}}\left(1+\frac{3}{8}X(z)\tau_O\right)\, .
\ee
The phase $\Phi_O(\tau_O)$ can be obtained by direct integration
of Eq.~\eqref{fOX}.
Alternatively, combining Eqs.~\eqref{fOX}, \eqref{fvstau} and \eqref{f} we find the
relation between the time to coalescence at the source and at the observer locations at linear order in $X(z)$:
\be
\label{tauX}
\tau_S=\frac{\tau_O}{1+\bar z} \Big[1-X(z)\tau_O \Big]\, .
\ee
Since the phase is constant $\Phi_O(\tau_O(\tau_S))=\Phi_S(\tau_S)$,
using Eq.~\eqref{OmS} and relating $\tau_S$ to $\tau_O$ with Eq.~\eqref{tauX}
we get
\be
\label{dphiO}
\Phi_O(\tau_O)=
-2\left(\frac{\tau_O}{5G\Mc} \right)^{\frac{5}{8}}\left(1-\frac 58X(z)\tau_O\right)+\Phi_c\, .
\ee
The acceleration (or deceleration) of the universe during the time of observation of the GW generates therefore an additional contribution to the observed frequency and the observed phase with a different time dependence. Note that if the expansion of the universe is constant in time, $H_S=H_0(1+\bar z)$ and as expected $X(z)$ exactly vanishes. 

\subsection{Perturbed universe}
\label{sec:pert}

Let us now add the effect of the redshift perturbations $\delta z$. We go back to Eq.~\eqref{dgO}, that we have to solve including also the redshift perturbations coming from the matter perturbations in Eq.~\eqref{dez}, so that at first order 
\be
\Big[1+z(t'_O)\Big]^{-1}=\frac{1-\delta z(t'_O)}{1+\bar z(t'_O)}
\ee
where from Eq.~\eqref{dez}
\begin{align}
\delta z(t'_O)= &\psi(t'_O)-\psi(t'_S) +\big[\bv(t'_S)-\bv(t'_O)\big]\cdot \bn\\
 &-\int_{t'_S}^{t'_O}  dt'(\dot{\phi}+\dot{\psi})\, .\nonumber
\end{align}
Note that here $t'_S$ is a function of $t'_O$. We can expand $\delta z(t'_O)$ around $t_O$ and $t_S$:
\begin{widetext}
\bea
\delta z(t'_O)&=& \psi(t_O) +\dot{\psi}(t_O)(t'_O-t_O)-\psi(t_S) -\dot{\psi}(t_S)(t'_S-t_S)
-\big[\bv(t_O)+\dot{\bv}(t_O)(t'_O-t_O)\big]\cdot\bn \nonumber\\ 
&&+\big[\bv(t_S)+\dot{\bv}(t_S)(t'_S-t_S)\big]\cdot\bn
-\int_{t_S}^{t_O}  dt'(\dot{\phi}+\dot{\psi}) -\big[\dot{\phi}(t_O)+\dot{\psi}(t_O)\big](t'_O-t_O)
+\big[\dot{\phi}(t_S)+\dot{\psi}(t_S)\big](t'_S-t_S)\nonumber\\
&=& \delta z(t_O) +\left(\frac{\dot{\bv}_S\cdot\bn}{1+\bar z}- \dot{\bv}_O\cdot\bn + \frac{\dot{\phi}_S}{1+\bar z}-\dot{\phi}_O\right)(t'_O-t_O)\, . \label{dz}
\eea
Inserting Eq.~\eqref{dz} into Eq.~\eqref{dgO} and integrating we obtain
\bea
f(\tau_O)&=&\frac{1}{\pi}\left(\frac{5}{256\,\tau_O}\right)^{3/8}\left(G \Mc \right)^{-5/8}\left[1+\frac{3}{8}X(z)\tau_O 
+\frac{3}{16}\left(\frac{\dot{\bv}_S\cdot\bn}{1+\bar z}- \dot{\bv}_O\cdot\bn + \frac{\dot{\phi}_S}{1+\bar z}-\dot{\phi}_O\right) \tau_O\right] \\
&\equiv& \frac{1}{\pi}\left(\frac{5}{256\,\tau_O}\right)^{3/8}\left(G \Mc \right)^{-5/8}\left[1+\frac{3}{8}Y(z)\tau_O \right]\,,\nonumber
\label{fOY}
\eea
\end{widetext}
where we have defined the variable $Y(z)$ which accounts for both the effect due to the background acceleration of the universe and for the one due to perturbations: 
\bea\label{Y}
Y(z)\equiv X(z)+\frac{1}{2}\left(\frac{\dot{\bv}_S\cdot\bn}{1+\bar z}- \dot{\bv}_O\cdot\bn + \frac{\dot{\phi}_S}{1+\bar z}-\dot{\phi}_O\right) \,.
\eea
As before we can find the relation between $\tau_S$ and $\tau_O$
\be
\tau_S= \frac{\tau_O}{1+\bar z} \left[1- Y(z)\tau_O\right]\,.
\label{tauY}
\ee
The phase at the observer is given by
\be
\Phi_O(\tau_O)=
-2 \left(\frac{\tau_O}{5G\Mc} \right)^{5/8} \left(1-\frac{5}{8}\,Y(z)\tau_O\right)+\Phi_c\,.\label{PhiO}
\ee
The phase can be directly inserted into the GW amplitudes at the observer Eqs.~(\ref{hO},\,\ref{hOcross}), which therefore acquire an explicit dependence on 
redshift through $Y(z)$.
Moreover, the common amplitude of the two polarisations of the GW $h_c(\tau_O)$, when written in terms of the time to coalescence, is also affected by the evolution of the background plus perturbations, and
becomes:
\be
\label{hctau}
h_c(\tau_O)=
\frac{(G \mathcal{M}_c(z))^{5/4}}{d_L(z)}\left(\frac{5}{\tau_O}\right)^{1/4}\left(1+\frac{1}{4}\,Y(z)\tau_O\right)\,.
\ee
We emphasise that the $Y(z)$ effect is obtained when expressing everything in
terms of $\tau_O$. The amplitude and the phase of the waveform would take the standard form if
expressed in terms of $f_O$, since the frequency is simply redshifted as in
Eq.~\eqref{f}, while a time interval depends on the variation of the expansion of the universe and on the variation of the perturbations as in \eqref{tauY}. 

Eqs.~\eqref{PhiO} and~\eqref{hctau} show how the peculiar acceleration of the binary and of the observer and the time variation of the gravitational potential modify the phase and the amplitude of the GW. Note that the amplitude is also affected by the fluctuations in the luminosity distance $d_L$ due to matter inhomogeneities~\cite{distance, Hui:2005nm}. 

\vspace{0.29cm}
\section{Modifications of the waveform}
\label{se:waveform}

So far we have calculated how the background expansion of the universe and the
redshift perturbations affect the frequency, the phase and the amplitude of the chirp signal in
time. In order to quantify whether
these effects are detectable, and make contact with standard data analysis techniques, here we consider the Fourier
transform of the GW-form. This is calculated using the stationary phase approximation, and at leading
order in the amplitude is (see e.g.~\cite{michele}):
\bea
\tilde h_+(f) &=&A(f) \frac{1+\cos^2\imath}{2} \, \exp(i\Psi(f))\,,  \\
\tilde h_\times(f) &=&A(f)  \cos\imath \,\,\exp\left(i\Psi(f) +\frac{\pi}{2}\right)\,.
\eea
Written in terms of $\tau_O(f_O)$, the phase $\Psi(f)$ at leading order and for non-spinning inspiralling
binaries is
($t_c$ is the time of coalescence at the observer)
\begin{widetext}
\bea\label{Psi+}
\Psi(f) &=& 2\pi f t_c -\frac{\pi}{4} -2\pi f \tau_O(f)-\Phi_O(\tau_O(f)) \nonumber\\
&=& 2\pi f t_c -\frac{\pi}{4} -\Phi_c + \frac{3}{128} \left(\pi G\mathcal{M}_c\right)^{-5/3}
\frac{1}{f^{5/3}} -\frac{25}{32768\,\pi} \left(\pi G\mathcal{M}_c\right)^{-10/3} \frac{Y(z)}{f^{13/3}} \,,
\eea
where in the second equality we have used the inverse of Eq.~(\ref{fOY}) and Eq.~(\ref{PhiO})
to include the leading order effect in $Y(z)$. In terms of its frequency dependence, this effect is formally a $-4$PN effect. The PN expansion parameter is indeed $x\equiv (\pi G\mathcal{M} f_O)^{2/3}\sim V^2$, where $V$ denotes the velocity of the inspiralling masses around the center of mass of the binary (which should not be confused with the peculiar velocity of the binary with respect to the Hubble flow ${\bf v}_S$), and the Newtonian term is proportional to $x^{-5/2}$. 
The amplitude of the wave gets multiplied by the (inverse of the) second time
derivative of the phase, leading to 
\bea\label{Af}
A(f)=\sqrt{\frac{\pi}{2}}h_c(f_O) \left( \frac{{\rm d}^2 \Phi_O}{{\rm d}\tau_O^2}\right)^{-1/2}
=\sqrt{\frac{5}{24 \pi^{4/3}}} \,  \frac{(G\mathcal{M}_c)^{5/6}}{d_L(z)}  \, \frac{1}{f^{7/6}} \, \left[ 1-\frac{5 \,(G\mathcal{M}_c)^{-5/3}}{384\pi^{8/3}} \, \frac{Y(z)}{f^{8/3}} \right]\,.
\eea
This effect on the amplitude has been derived here for the first time. 
\end{widetext}

\section{Estimate of the effect}
\label{se:mismatch}

\subsection{Comparison between the contributions from the background expansion and the redshift perturbations}

In the previous section we have shown that the background expansion of the universe and the presence of perturbations in the redshift affect the waveform introducing a frequency-dependent term with amplitude $Y(z)$, to which they contribute with different strength, as given in Eq.~\eqref{Y}.
From Eq.~\eqref{Y} we see that the contribution to $Y(z)$ from the redshift perturbations depends on the time derivative of the peculiar velocity $\bv_S$ and of the potential $\phi$ at the source~\footnote{Note that the quantities at the observer also play a role, but we do not consider them here: the acceleration of the Earth $\dot{\bf v}_O$ is already taken into account in GW observations, and the time variation of the gravitational potential $\dot\phi_O$ is small.}. The potential $\phi$ varies on a cosmological timescale: one has therefore $\dot\phi\sim H_0\phi \sim 10^{-5} H_0$, while $X(z)\simeq H_0$. The  contribution to $Y(z)$ from the gravitational potential is therefore smaller than the one from the background expansion by a factor of $10^{-5}$, and we neglect it in the following. 

The effect from the variation of the peculiar velocity of the source is instead stronger, as we will now show. Assuming that the source is within a galaxy which is itself within a cluster, the peculiar velocity can be split into three contributions:
\begin{enumerate}
\item the velocity of the cluster (which contains the binary) with respect to the CMB frame
\item the velocity of the galaxy inside the cluster
\item the velocity of the centre of mass of the binary inside the galaxy
\end{enumerate}

\noindent{\bf Contribution 1}:
The first velocity contribution is ascribable to the cosmological linear velocity perturbation $\bv_S^{\rm cosmo}$ and it varies on a cosmological timescale:
like for the variation of the scalar potential, one has $\dot \bv_S^{\rm cosmo} \sim H_0 \bv_S^{\rm cosmo} \sim 10^{-3} H_0$. The contribution of this velocity to $Y(z)$ is therefore always smaller than the one of $X(z)$ and we neglect it. 

The second and third contributions, on the other hand, can vary on a timescale smaller 
than the cosmological timescale, and they can therefore be bigger than the background expansion effect represented by $X(z)$.
In order to make the comparison with $X(z)$, we define the variable (c.f.~Eq.~\eqref{Y})
\be\label{F}
\frac{F(z)}{H_0}\equiv\frac{1}{2}\frac{\dot{\bv}_S\cdot\bn}{(1+\bar z)H_0}\,,
\ee 
to be compared with $X(z)/H_0$. We estimate the two remaining contributions to the peculiar velocity of the source $\bv_S$ as follows:

\noindent{\bf Contribution 2}: For a virialised cluster of mass $M$ one has, at distance $r$ from the centre, 
\be
\frac{5 G M}{3r}\simeq \frac{1}{2}v_S^2\, .
\ee
Combining this with Newton's law we find for the peculiar acceleration
\be
\dot{v}_S\simeq \frac{3}{10} \frac{v_S^2}{r}\, .
\ee

\vspace{0.5cm}

\noindent{\bf Contribution 3} : For a circular motion of the binary around the centre of the galaxy one has
\be
\dot{v}_S=\frac{v_S^2}{r}\, .
\ee

The function $F(z)$ can then be written as
\be
\frac{F(z)}{H_0}=\frac{\alpha}{2}\frac{v_S^2}{r}\frac{\mathbf{e}\cdot\bn}{H_0(1+\bar z)}\, ,
\ee
where $\alpha=3/10$ describes the galaxy's acceleration in the cluster while $\alpha=1$ describes the binary's acceleration inside the galaxy, and $\mathbf{e}$ denotes the direction of the acceleration. The total effect is the sum of the two relevant contributions. Defining $\epsilon$ as
\be
\label{epsilon}
\epsilon=\alpha\left(\frac{v_S}{100\,{\rm km/s}} \right)^2 \frac{10\,{\rm kpc}}{r}\,\mathbf{e}\cdot\bn\, ,
\ee
we obtain
\be
\label{Fbin}
\frac{F(z)}{H_0}=2.4\times 10^{-2} \frac{\epsilon}{1+\bar z}\, .
\ee
The amplitude of the effect depends therefore on three quantities: 1) the amplitude of the velocity $v_S$; 2) the distance of the binary from the centre of the galaxy and the distance of the galaxy from the centre of the cluster; and 3) the orientation of the acceleration with respect to the direction of observation. In the following we consider various concrete examples for the value of the parameter $\epsilon$ representing these contributions.

\begin{figure}[t]
\centering
\includegraphics[width=\columnwidth]{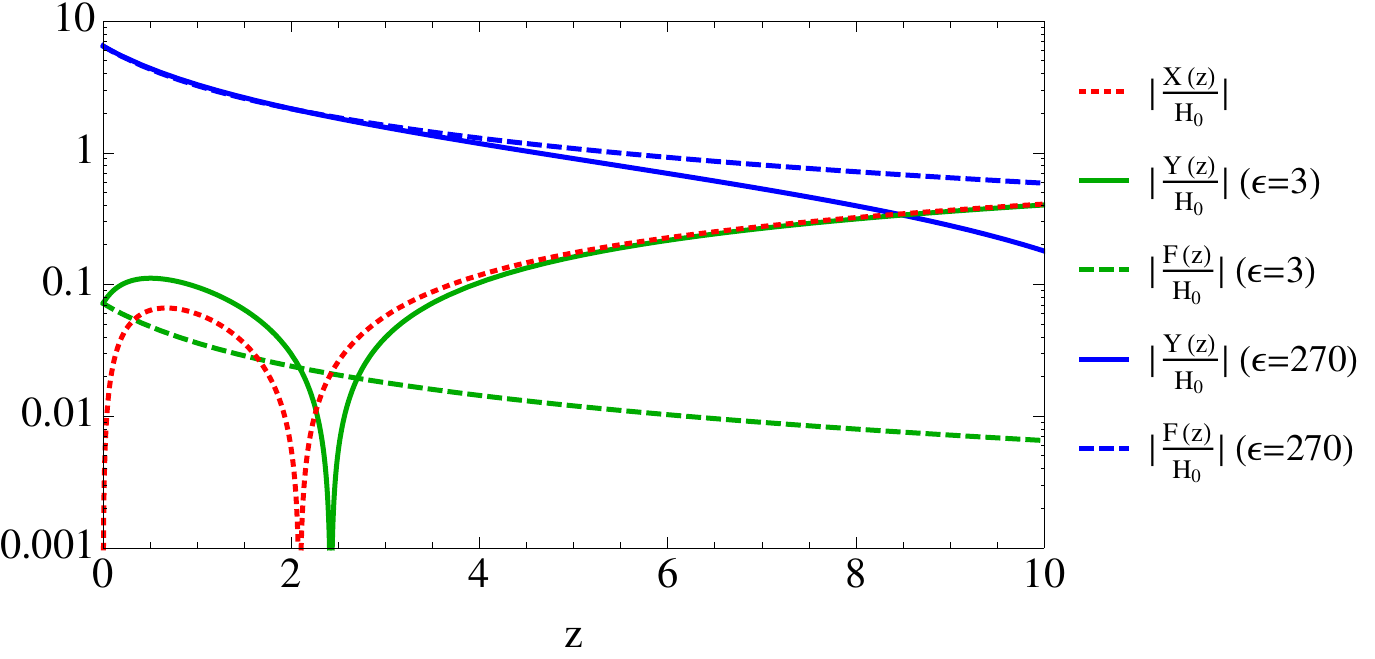}
\caption{\label{fig:Fz}
As a function of redshift, the red dotted line shows
the absolute value of the effect of the time variation of
the background expansion on the GW phasing $X(z)$ (Eq.~\eqref{eq:Xz}) relative to $H_0$. The dashed lines show the GW phasing variation due to the peculiar acceleration
of the binary $F(z)$ (Eq.~\eqref{Fbin}) relative to $H_0$. The solid lines show the absolute value of the sum of the two effects $Y(z)$ (Eq.~\eqref{Y}) relative to $H_0$. Blue lines are for $\epsilon = 270$ and green lines for $\epsilon = 3$.
Note that the effect due to the expansion becomes negative at large enough redshift,
and therefore the two contributions eventually partially compensate each other.
For $\epsilon = 270$ this happens at very high redshift since the effect from the peculiar acceleration always dominates over the effect from the expansion up to $z = 10$. For $\epsilon = 3$ the two effects cancel at $z\approx 2.5$ and are comparable for lower redshift, while at higher redshift the expansion effect dominates over the one of the peculiar acceleration.}
\end{figure}

In figure~\ref{fig:Fz} we compare the effect of the background acceleration $X(z)/H_0$
with the one from the peculiar acceleration of the binary $F(z)/H_0$. For simplicity we account only for the peculiar acceleration of the galaxy with respect to the cluster (contribution 2), and we neglect the acceleration of the binary inside the galaxy (contribution 1), which is generally subdominant. We consider two cases: a galaxy with high peculiar velocity $v_S = 3000\, {\rm km/s}$ very close to
the center of the cluster $r = 10\, {\rm kpc}$, giving $\epsilon=270$; and a binary at the edge of the cluster $r = 100\, {\rm kpc}$ with average velocity $v_S = 1000\, {\rm km/s}$, giving $\epsilon=3$. In the first case, it appears that the peculiar acceleration of the binary dominates the contribution due to the universe acceleration for redshift smaller than about 7. In the second case, the two contributions are of the same order of magnitude at small redshift and the background acceleration dominates over the peculiar acceleration at high redshift.

We have shown that, contrary to $X(z)$ which only depends on redshift and cosmological parameters, $F(z)$ further depends on the unknown peculiar acceleration of the binary along the line of sight $\dot \bv_S\cdot {\bf n}$.
This dependence on an unknown extra parameter contaminates the determination of the redshift that would have been possible if only $X(z)$ would be present in Eqs.~\eqref{Psi+} and \eqref{Af}, instead of the total effect $Y(z)$. 
As a consequence, we expect the contribution from the peculiar acceleration of
the binary to seriously degrade the possibility of using the binaries as standard candles without redshift counterpart, analysed in \cite{Nishizawa}. Note that one possibility to get rid of the acceleration effect would be to average the signal from a large number of sources, as mentioned in~\cite{Seto:2001qf}. Since in average the sources are moving with the Hubble flow, the peculiar acceleration over a sufficiently large number of sources is expected to vanish, leaving only the impact from the acceleration of the background. We highlight however that such an average is non-trivial to perform: one needs first to determine the luminosity distance of the binaries in order to average over binaries that are at the same distance, and therefore at the same redshift. Since the luminosity distance is itself affected by inhomogeneities in the matter distribution, e.g. through the effect of gravitational lensing, such an average will be affected by these inhomogeneities (for example average will be taken over binaries at different redshifts but happening to have the same luminosity distance because of different
lensing effect on one binary than on the other). The second difficulty is that the amplitude of the acceleration effect depends not only on $Y(z)$ but also on the redshifted chirp mass of the binary (see eq.~\eqref{Psi+}). We can therefore not simply average over all binaries at the same distance, but we need first to determine $\mathcal{M}_c$ and $Y(z)$ and then average over $Y(z)$. These two non-trivial aspects of the average are expected to contaminate the determination of the mean redshift.

\subsection{Estimate of the phase shift}
\label{estimate}

Motivated by the fact that interferometric detectors of GWs are particularly sensitive to the GW phase,
let us use Eq.~\eqref{PhiO} to estimate the phase shift $\Delta\Phi_Y$ due to the total acceleration effect $Y(z)$ during the observation of a typical binary.
Let us assume that we observe the binary for a time interval $\Delta t=t_{\rm max}-t_{\rm min}=\tau_{O\,{\rm max}}-\tau_{O\,{\rm min}}$, thus
having
\be\label{DePhi}
\Delta\Phi_Y=\frac{5}{4} \left(5G\Mc \right)^{-5/8} Y(z) \, \Big[\tau_{O\,{\rm max}}^{13/8} -\tau_{O\,{\rm min}}^{13/8}\Big]\,.
\ee
The time to coalescence in terms of the chirp mass and frequency of the binary can be found for example in Eq.~(4.21) of \cite{michele} (we normalise it here for a typical binary that would be observable by LISA, having $m_1=m_2=5\cdot 10^3M_\odot$):
\be
\tau_{O\,{\rm max}}\simeq 1.4 \, {\rm year} \left(\frac{5\cdot 10^3 M_\odot}{\mathcal M_c(z)}\right)^{5/3} \left(\frac{10^{-3} {\rm Hz}}{f_O}\right)^{8/3}\,.
\ee
If we can observe the binary up to coalescence time, so that $\tau_{O\,{\rm min}}=0$, from Eq.~\eqref{DePhi} one has ($h$ comes from the Hubble factor today)
\begin{multline}\label{DePhiCoal}
\Delta\Phi_{Y,\rm coal}\simeq 3.96 \cdot 10^{-5} h \\
\times \frac{Y(z)}{H_0}\left(\frac{5\cdot 10^3 M_\odot}{\mathcal M_c(z)}\right)^{\frac{10}{3}} \left(\frac{10^{-3} {\rm Hz}}{f_O}\right)^{\frac{13}{3}}\,.
\end{multline}
We see that the shift in the phase is larger if we observe binaries at low frequency,
with small (redshifted) chirp mass. 

However, decreasing the mass of the binary makes the signal
exit the detector band long before it merges.
Let us therefore rewrite Eq.~\eqref{DePhi} under the assumption that we observe the binary for a time interval
$\Delta t$ smaller than the time to coalescence.
In this case one has, in the limit $\Delta t\ll \tau_{O\,{\rm max}}$,
\bea\label{DePhiDet}
\Delta\Phi_{Y,\rm \Delta t} &\simeq& \frac{65}{32} \left(5G\Mc \right)^{-5/8} Y(z) \, \tau_{O\,{\rm max}}^{5/8}  \Delta t \nonumber\\
& \simeq & 0.1 \, h\,
\frac{Y(z)}{H_0} \left(\frac{50 M_\odot}{\mathcal M_c(z)}\right)^{\frac{5}{3}} \left(\frac{10^{-3} {\rm Hz}}{f_O}\right)^{\frac{5}{3}} \frac{\Delta t}{{\rm year}}~~~~~
\eea
where we have normalised the redshifted chirp mass to 50 $M_\odot$. 

From the above equations \eqref{DePhiCoal} and \eqref{DePhiDet},
it appears that a significant dephasing can be achieved for
those binary systems which are relevant for LISA, provided that the (redshifted) chirp mass of the
binary system is low enough and that the amplitude of the effect $Y(z)$ is large enough. In particular, the range of binary masses for which we expect the acceleration effect to influence the measurement corresponds to a few tens of solar masses (for almost equal mass binaries). 
The above equations also show that a significant dephasing can never be achieved in the frequency range accessible to Earth based interferometers: the acceleration effect is relevant neither for advanced LIGO and Virgo, nor for future detectors as the Einstein Telescope. 

\begin{figure}[t]
\includegraphics[width=1\linewidth]{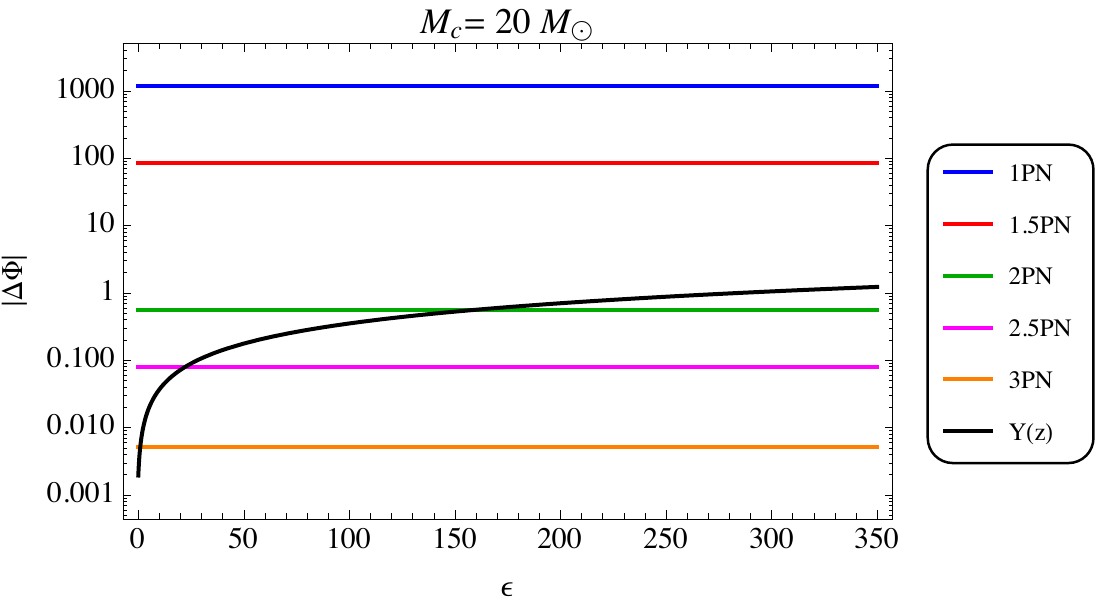}
\caption{Absolute value of the total phase shift as a function of $\epsilon$ due to the peculiar acceleration and the expansion of the universe, compared to the phase shift one expects from the first five PN terms, for an equal mass binary with $\mathcal{M}_c=20\,M_{\odot}$, at redshift $z=0.1$, that enters the LISA detector at $f_{\rm min}=0.004$ Hz and remains in band for five years. The amplitude of the effect due solely to the expansion of the universe is given by the value of the black curve taken at $\epsilon=0$.}
\label{fig:PN}
\end{figure}

In order to provide a comparison of the total acceleration effect found here
with quantities known in the literature, in Fig.~\ref{fig:PN} we present the total phase shift due to the combined effect of peculiar acceleration and the expansion of the universe together with the phase shifts induced by the first three standard PN terms according to the formula
\bea
\label{Psi3.5PN}
\lefteqn{\Psi(f) = 2\pi f t_c -\frac{\pi}{4} -\Phi_c} \\
& &+\frac{3}{128\,\eta\, x^{5/2}}\left(1+\sum_{i=1}^7a_i(\eta) x^{i/2} -\frac{25}{768}\frac{G \mathcal{M}_c Y(z)}{\eta^{8/5}\,x^4}\right)\,, \nonumber
\eea
where the explicit values for the $a_i(\eta)$ are given in~\cite{Mishra:2010tp}
(note that $a_{5,6}$ have a logarithmic dependence on $x$, here understood
as a function of $f_O$). We consider an equal mass binary with $\mathcal{M}_c=20\,M_{\odot}$, at redshift $z=0.1$, that enters the LISA detector at $f_{\rm min}=0.004$ Hz and remains in band for five years (c.f. section \ref{LISAbin} for a discussion on how to determine the final frequency after five years of observation). Under these conditions, we find that the phase shift lies between those due to the 2.5PN and 2PN terms, for $\epsilon\gtrsim 20$: it is therefore in principle detectable\footnote{Note that, as stressed above, Eq.~\eqref{Psi+} shows that the acceleration effect has the same {\it frequency dependence} of a -4PN effect, i.e. $f^{-13/3}$. However, we have demonstrated here that, since the amplitude of the acceleration effect is small, in terms of phase shift and detectability it is comparable to at best a 2PN term.}.

In the following sections, we analyse more quantitatively the total acceleration
effect concentrating on the case of LISA.
As we have seen the effect is higher for lower masses at fixed frequency. 
However, when the GW signal is deeply in the inspiral regime it is almost monochromatic (almost negligible chirping): we expect that the phase shift evaluated above 
can be re-absorbed by a slight shift of the values of the binary parameters, like the symmetric mass ratio $\eta$ and the time of coalescence $t_c$. This will be shown by a more precise computation in the next section; here we provide a semi-qualitative argument to understand how this happens. Denoting by $\Delta\Phi_{\rm N}$ the GW-phase accumulated during the
observation at Newtonian order (the main contribution), a variation of the source chirp mass (symmetric mass ratio)
$\delta M_c$ ($\delta\eta$) can induce a de-phasing $\delta\Phi$ of the order \\

\begin{align}
&|\delta\Phi_{\eta={\rm const}}|\simeq \frac 58 \Delta\Phi_{\rm N} \frac{\delta M_c}{M_c} \label{dePHIetaconst} \\
&|\delta\Phi_{M_c={\rm const}}|\simeq 0.4 \, |\Phi_{\rm N}^{\frac35}(f_{\rm max})-\Phi_{\rm N}^{\frac35}(f_{\rm min})| \,
\frac{\delta\eta}{\eta^{7/5}}\,. \label{dePHIMcconst}
\end{align}
For the typical objects we are interested in, taking as example the same parameters adopted in Fig.~\ref{fig:PN}, the dephasing at Newtonian order is $|\Delta\Phi_{\rm N}|\simeq 10^6$. 
Given this value, phase shifts of $O(1)$ (which are typically detectable) can be produced by $\delta M_c/M_c\sim 10^{-6}$ and
$\delta\eta/\eta\sim 3\cdot 10^{-3}$ (for $\eta=0.25$).
We expect then that the impact of the peculiar acceleration can be
absorbed by a tiny shift in the intrinsic parameters. We now proceed to quantify such \emph{bias} on the intrinsic parameters in order to estimate its impact. In particular, since the waveform is extremely sensitive
to $M_c$, we expect the stronger bias to occur on $\eta$.

\subsection{The mismatch}

In order to have a more quantitative estimate of the importance of the total acceleration effect $Y(z)$
and its detectability, we quantify the difference between GW forms with and without the effect (respectively, injection and template) by introducing the scalar product as the noise-weighted frequency
\emph{overlap} between two waveforms $h_{1,2}$ \cite{Cutler:1994ys}
\be
\langle h_1|h_2\rangle\equiv 2\int_0^\infty
\frac{\tilde h_1(f)\tilde h^*_2(f)+\tilde h_2(f)\tilde h_1^*(f)}{S_n(f)}\,df\, ,
\ee
where $S_n(f)$ is the one-sided noise spectral density of the GW detector. The above scalar product  
allows to define a norm of a function $h_1$ as
$||h_1||\equiv (\langle h_1 |h_1\rangle)^{1/2}$.

The fitting factor $FF$ between two waveforms $h_{1,2}$
(for concreteness we can think of them as respectively the injection/data
and the template) is defined as the normalized overlap
maximised over search parameters \cite{Apostolatos:1995pj} 
\be
\label{ff}
FF\equiv\mathop{\rm Max}_{\Delta t_c,\Delta\Phi_c,\Delta M_c,\Delta\eta}
\frac{\langle h_1|h_2\rangle}{||h_1||\,||h_2||}\,.
\ee
Here we cover the simplified case of a spin-less circular binary inspiral,
so that the waveform overlap is only sensitive to phase and time,
beside binary constituent individual masses.
In this simplified case geometric factors due to orientation as well as distance of the
source from the observer boil down to a mere rescale of the waveform, thus not affecting
the normalised overlap.\footnote{We neglect the amplitude modulation
due to the changing sky position during the observation time, assuming
the GW data are suitably adjusted to take into account this annual modulation.}
Computationally the maximisation over $\Delta\Phi_c$ 
can be done
analytically \cite{Allen:2005fk} by introducing
\bea
\mathcal{O}&\equiv&
\mathop{\rm Max}_{\Delta t_c\Delta\Phi_c}\frac{\langle h_1|h_2\rangle}{||h_1||\,||h_2||} \nonumber \\
\label{eq:maxOver}
&=& \mathop{\rm Max}_{\Delta t_c}
\frac{4\left|\int_0^\infty{\tilde h_1(f) \tilde h_2^*(f) e^{2\pi i ft}/S_n(f)}\right|}{||h_1|||h_2||}
\eea
at the computationally inexpensive price of taking a Fourier transform of the
waveform overlap, thus having 
\begin{equation}
\label{eq:ff}
FF=\mathop{\rm Max}_{\Delta M_c,\Delta\eta}\mathcal{O}\,.
\end{equation}

The maximisation is done via a Monte Carlo code implementing the
simulated annealing algorithm\footnote{The simulated annealing algorithm
is particulararly suited for getting quickly to the maximum of the distribution
to which we are interested in, rather than its profile.}
 \cite{Kirkpatrick:1983zz,Cerny:1982rt},
searching over the two-dimensional $M_c-\eta$ plane
for the best fit waveform which is chosen by maximising over $t_c,M_c,\eta$ as
per \eqref{eq:maxOver} and \eqref{eq:ff}. 

The mismatch is defined as \cite{Buonanno:2009zt}
\be m\equiv 1-FF\,. \ee 
For the computation of the overlap we use the waveform amplitude at the lowest post-Newtonian
order and for the phase we use the highest order available analytically, i.e.~3.5 PN, see Eq.~\eqref{Psi3.5PN}. For the acceleration effect
we keep only the leading effectively -4PN order.

We compute the mismatch $m$ over a set of injections simulating
physical signals with several values of masses, redshift and $\epsilon$ and
recovering them with signals not containing the acceleration effect, i.e.
waveform characterised by phase \eqref{Psi3.5PN} with $Y=0$. This reproduces the
situation in which real signals are searched for with templates
not taking into account the acceleration effect and it would allow us to determine how many signals
will be lost and/or how biased the astrophysical parameter reconstruction would
be.

\subsection{The total acceleration effect on binaries visible by LISA}
\label{LISAbin}

\begin{figure}[t]
\includegraphics[width=0.9\linewidth]{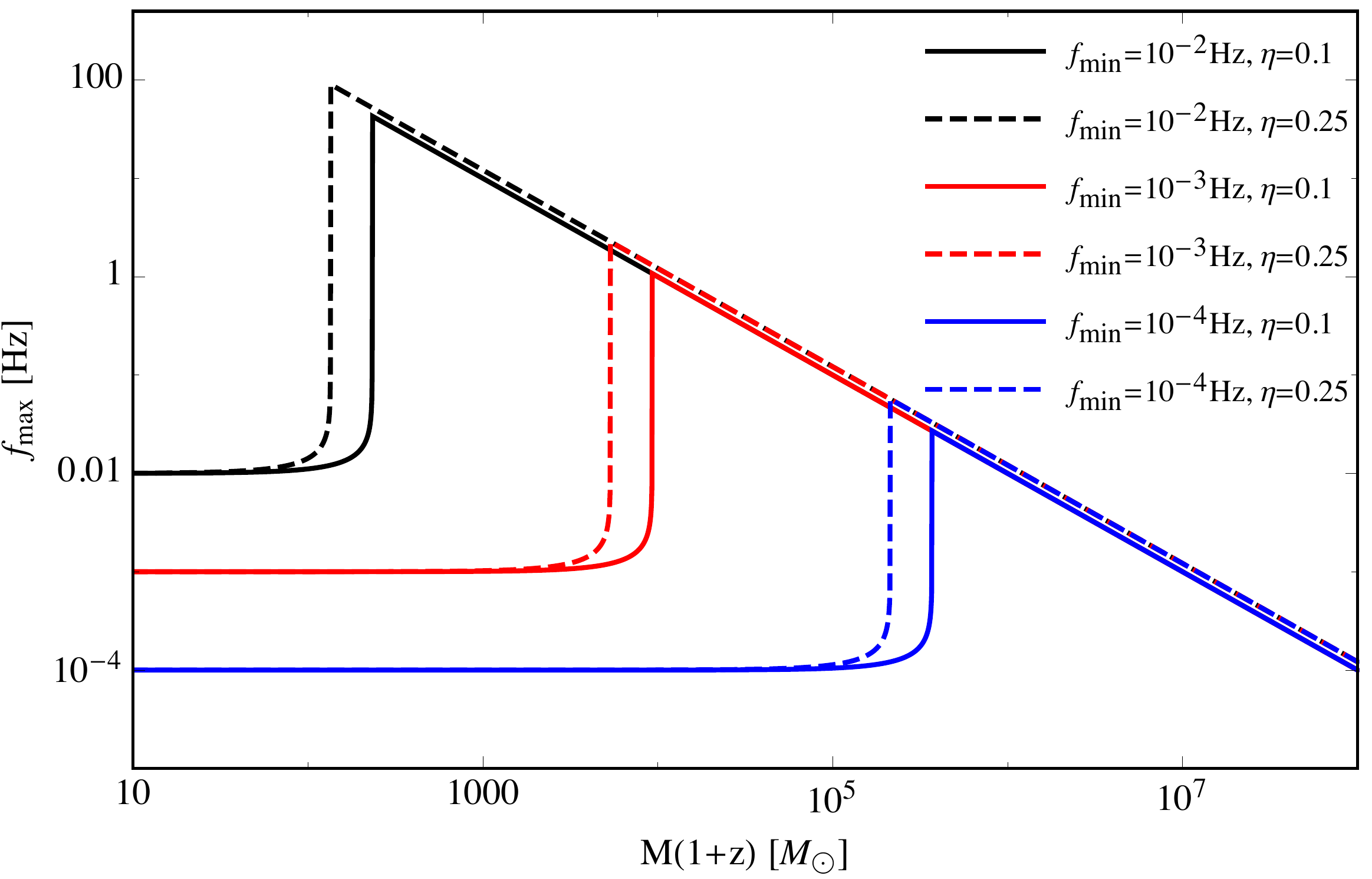}
\caption{$f_{\rm max}^{\rm LISA}$ from Eq.~\eqref{fmaxLISA} as a function of the redshifted total mass $\mathcal{M}(z)=(m_1+m_2)(1+z)$ for several values of $f_{\rm min}$ and $\eta$ and for $\Delta t=5$ years. Note the non-monotonic behavior: at first increasing
the mass makes $f^{\rm LISA}_{\rm max}$ larger because the phase chirps more, when
eventually $f_{\rm ISCO}$ comes in band $f^{\rm LISA}_{\rm max}$ decreases inversely
proportional to the total mass.}
\label{fiscovsfmax}
\end{figure}

\begin{figure*}[t]
\centering
\includegraphics[width=\linewidth]{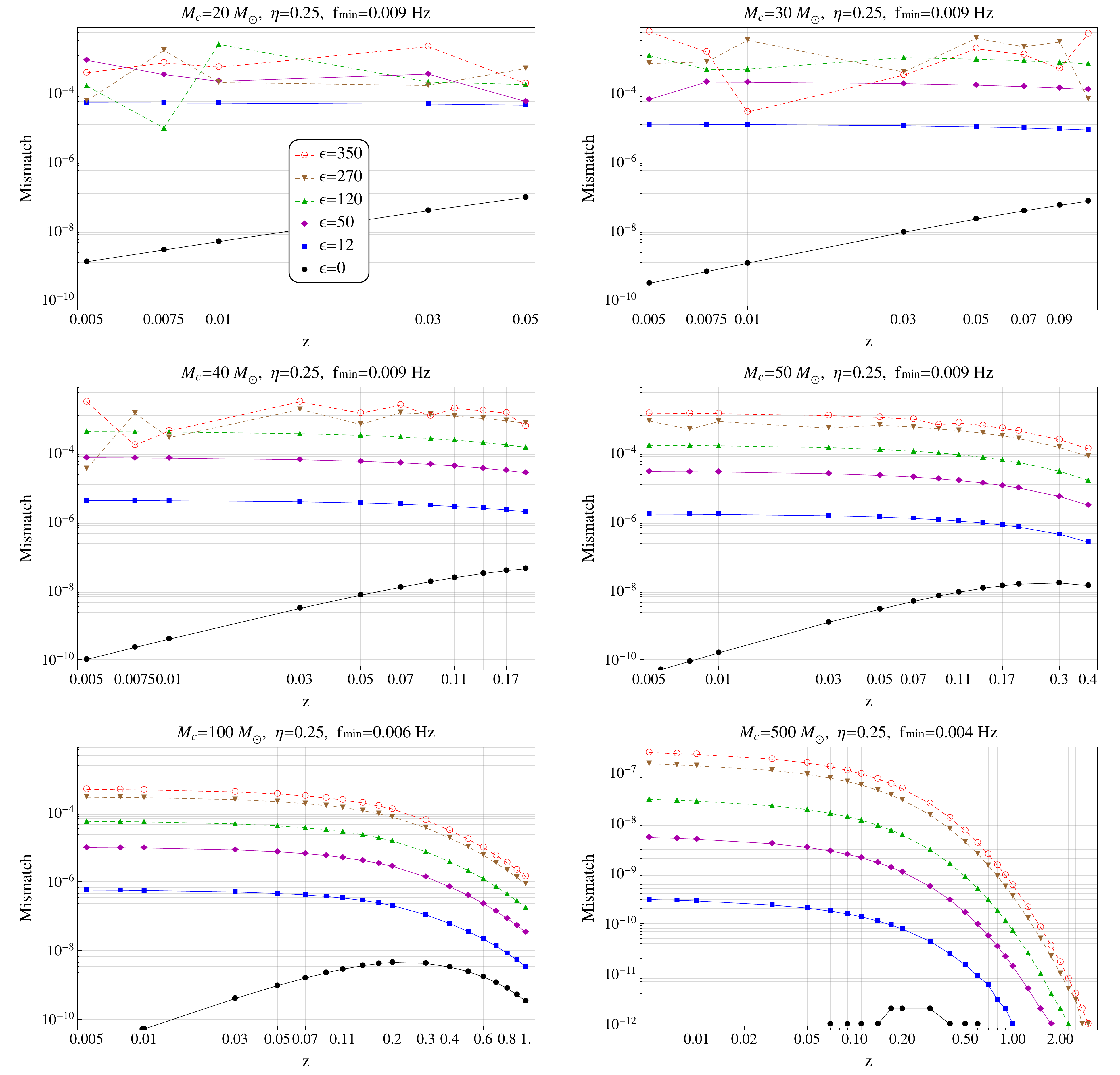}
\caption{Mismatch vs. redshift for $\eta=0.25$ and several values of $M_c$
at the source. The range in redshift for each $M_c$ has been determined by the requirement that the LISA SNR is larger or equal to 5. The minimal frequency for each mass value has been chosen within the interval given in Eq.~\eqref{fmin} in order to maximise at the same time the mismatch and the number of points in redshift, as discussed in the main text. For a large acceleration $\epsilon> 50$ and low masses $M_c< 50 \, M_\odot$ the result of the Monte Carlo code is noisy and the value of the mismatch has to be considered as an order of magnitude estimate: the corresponding lines are dashed (see discussion in the main text). Filled markers denote events for which $\eta$ is recovered within 1\% and $t_c$ is recovered within 60 seconds of their corresponding injected values, while empty markers denote events for which at least one of these two conditions is not satisfied.}
\label{fig:LISA}
\end{figure*}

\begin{figure}[t]
\centering
\includegraphics[width=\columnwidth]{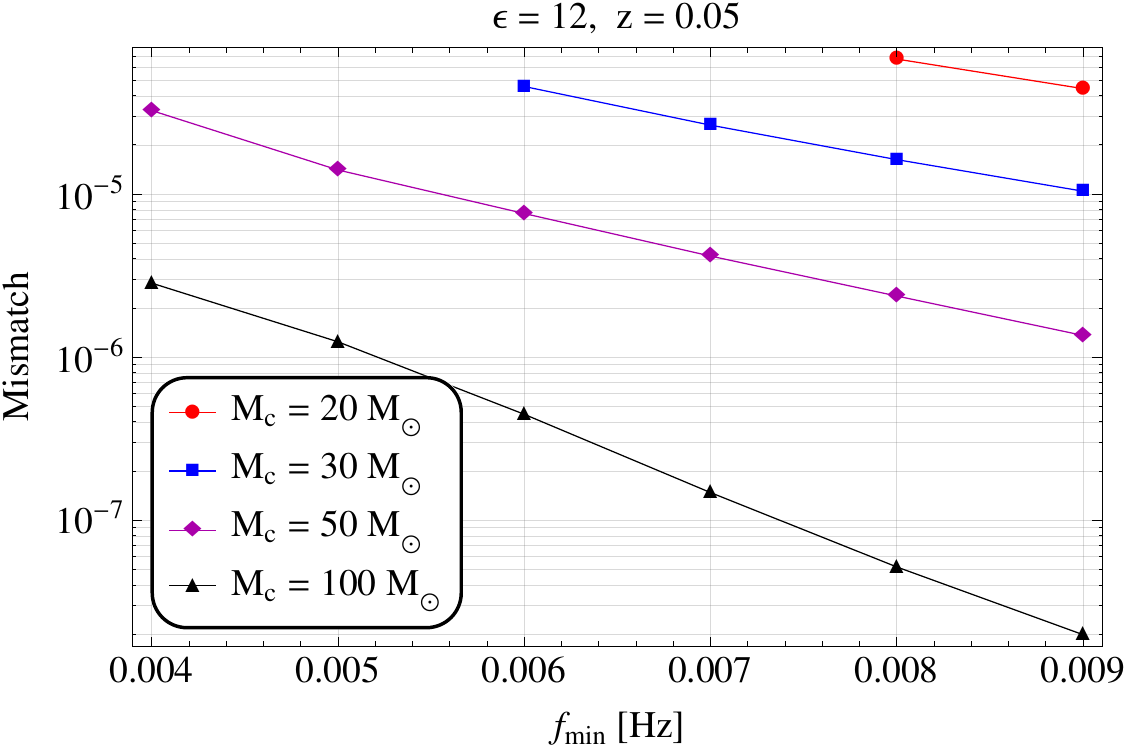}
\caption{Mismatch as a function of $f_{\rm min}$ for four values of the chirp masses at the source and fixed redshift.}
\label{fig:fmin}
\end{figure}

\begin{figure*}
\centering
\includegraphics[width=\linewidth]{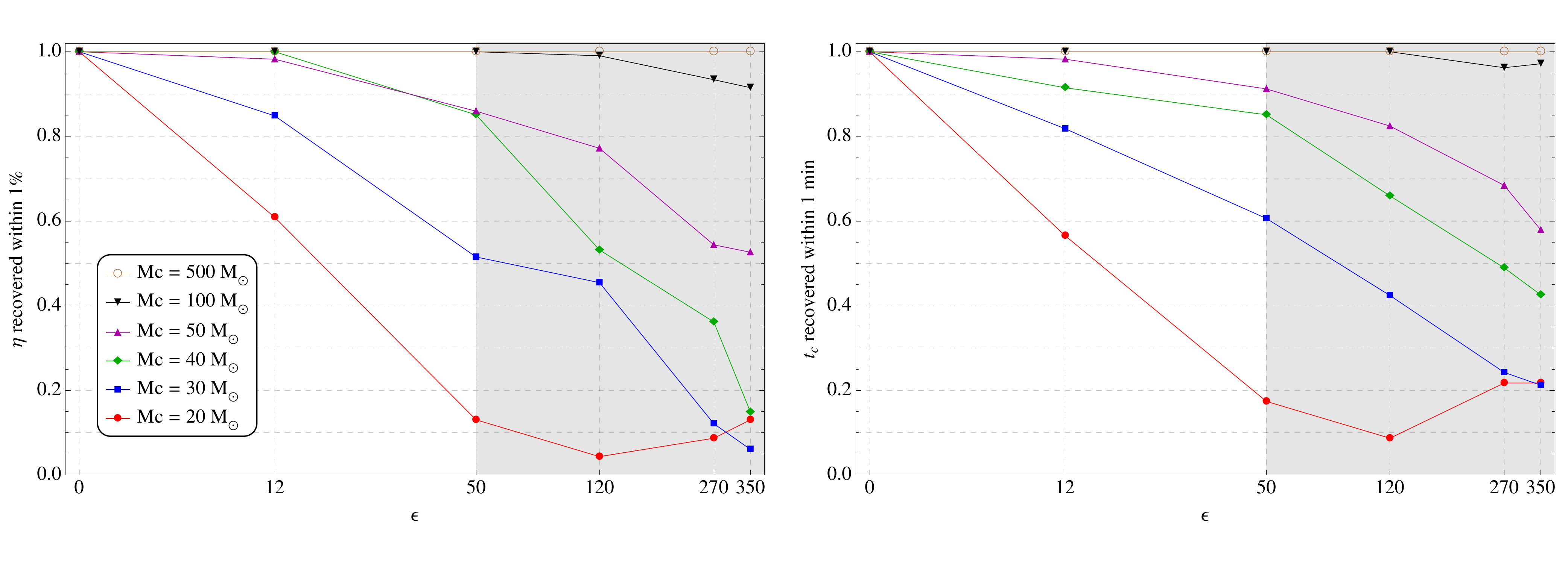}
\caption{Left panel: as a function of $\epsilon$, fraction of events for which the injected $\eta$ is recovered better than 1\% for several values of the masses. Right panel: as a function of $\epsilon$, fraction of events for which the injected $t_c$ is recovered within one minute for several values of the masses. For $\epsilon>50$ the results for low masses $M< 50 M_\odot$ can be affected by the Monte Carlo finite precision, as discussed in Sec.~\ref{LISAbin}: the regions where this can happen are grey shaded as a remainder.}
\label{fig:eta}
\end{figure*}

In this subsection we focus on LISA as the GW detector and determine the mismatch and the recovered binary parameters for different values of the binary masses and different strengths of the acceleration effect.

We consider the LISA configuration with 6 links (3 arms) (L6),
2 million km arm-length (A2), the ``expected'' LISA pathfinder low-frequency acceleration
noise (N2) and a mission duration of 5 years (M5): configuration id N2A2M5L6
according to the nomenclature of \cite{Klein:2015hvg}. 
The analytic fit to the LISA noise curve, including the white dwarfs confusion noise, is taken from \cite{Klein:2015hvg}: section II, Eqs.~(1) and (2).

The LISA low-frequency noise level has been recently tested by the LISA Pathfinder mission \cite{PATHFINDERweb} and, according to the first results, the expected noise is about six times
better than the original requirement for the Pathfinder, increasing to two orders of magnitude better at $f>60$ mHz \cite{PATHFINDERpaper}. The noise that we adopt in this analysis (N2) 
has been verified at frequencies $f > 1$ mHz, while at lower frequencies the situation is still
open: however one can optimistically forecast that the N2 noise level, if not a better
one, will be finally achieved over the whole frequency spectrum.
Concerning the arm-length, the effect of changing the LISA configuration does affect the mismatch since the noise spectral density
is shifted in frequency and the confusion noise due to white dwarfs binaries becomes relevant for longer arms \cite{Klein:2015hvg}.
Moreover, the signal to noise (SNR) ratio degrades for shorter arm configurations and shorter mission lifetime, so the range of possible detections might be strongly affected by this. We have chosen a LISA configuration of intermediate sensitivity to minimise these effects.

The main target sources for the LISA mission are super massive black hole binaries in the range $10^4 M_\odot < M < 10^7 M_\odot$.
However, LISA will also detect black hole binaries with masses of a few tens of solar masses, of the type of those seen by advanced LIGO 
\cite{Sesana:2016ljz, Vitale:2016rfr}.
Such low mass black hole binaries are the ones most affected by the acceleration
effect, c.f.~Eqs.~\eqref{DePhiCoal} and \eqref{DePhiDet}.
In what follows we consider six values of intrinsic chirp mass for the
black hole binaries that can possibly be detected by LISA:
$\mathcal M_c = 20, 30, 40, 50, 100, 500 M_\odot$, fixing for simplicity $\eta=0.25$.

The lower limit of integration to calculate the mismatch $m=1-FF$ 
is taken in the frequency region 
\be
4 \cdot 10^{-3} \,{\rm Hz}\leq f_{\rm min} \leq 9 \cdot 10^{-3}\,{\rm Hz}\,, 
\label{fmin}
\ee
which maximises the effect. The upper limit is set either by the maximum frequency attained by the binary (for which
we take twice the inner-most-stable-circular-orbit frequency as a proxy) or
by the frequency reached after the period of observation, which we set to the
foreseen duration of the mission: $\Delta t=5$ years.
More precisely we define
\begin{equation}
 f_{\rm max}^{\rm LISA} = {\rm min}[2 f_{\rm ISCO}, f(f_{\rm min},\Delta t)]\, ,
\label{fmaxLISA}
\end{equation}
where
\be
f_{\rm ISCO} \simeq 4.40\left(1+1.25 \eta+1.08\eta^2 \right) \left( \frac{M_\odot}{(1+z)M} \right) {\rm kHz}
\ee
is the frequency corresponding to the last stable orbit
\cite{Barausse:2009xi} and
\be
f(f_{\rm min},\Delta t)=\left[\frac{1}{f_{\rm min}^{8/3}} -
\frac{256 \pi^{\frac{8}{3}}}{5} \left(G\mathcal M_c(z) \right)^{\frac{5}{3}} \Delta t \right]^{-3/8}
\ee
is the instantaneous GW-frequency when a time $\Delta t$ has elapsed and the frequency
has evolved from $f_{\rm min}$ to $f$, as estimated by the Newtonian
order\footnote{We are only
considering the inspiral phase and will not take into account the merger and
ring-down phases, where the acceleration effect is negligible.}. In Fig.~\ref{fiscovsfmax}, we plot $f_{\rm max}$
as a function of the redshifted total mass for several values of $f_{\rm min}$ and $\eta$: it appears that, for the chosen range of masses, the ISCO is never reached. 

The lower mass black holes ($M \lesssim 10^2 M_\odot$) are expected to be the residual of stellar collapse and not to form the nuclei of galaxies, as it is instead expected for the more massive black holes ($M > 10^3 M_\odot$). Therefore for the binaries composed of two low-mass black holes, we expect two contributions to the acceleration: one from the velocity of the binary inside the galaxy and the second one from the velocity of the galaxy inside the cluster. For the more massive binaries on the other hand, which are situated at the galactic centre, we expect that only the velocity of the galaxy inside the cluster will be relevant.
Taking into account realistic velocities and distances, we have considered six different values of $\epsilon$. In the first four cases we assume that the effect is due only to the acceleration of the galaxy inside the cluster, i.e.~$\alpha=3/10$ in eq.~\eqref{epsilon}:  
\begin{itemize}
\item $\epsilon=0$, corresponding to $v_S = 0$: the effect of the peculiar acceleration is absent and the only contribution to $Y(z)$ comes from the acceleration of the universe, $X(z)$
\item $\epsilon=12$, corresponding to $v_S = 2000\, {\rm km/s}$ and $r = 100\, {\rm kpc}$
\item $\epsilon=120$, corresponding to $v_S = 2000\, {\rm km/s}$ and $r = 10\, {\rm kpc}$
\item $\epsilon=270$, corresponding to $v_S = 3000\, {\rm km/s}$ and $r = 10\, {\rm kpc}$
\end{itemize}
We have also considered two cases in which we add the acceleration of the galaxy inside the cluster and the acceleration of the binary inside the galaxy, assuming that they are perfectly aligned. We assume two corresponding values of $\epsilon$, an intermediate one and a very high one:
\begin{itemize}
\item $\epsilon=50$, corresponding to a galaxy velocity of $v_S= 2000\, {\rm km/s}$ at $r = 30\, {\rm kpc}$ ($\alpha=3/10$), and a binary velocity of  $v_S=250$\, km/s at $r=6.2$\, kpc ($\alpha=1$)
\item $\epsilon=350$, corresponding to a galaxy velocity of $v_S= 3000\, {\rm km/s}$ at $r = 10\, {\rm kpc}$ ($\alpha=3/10$), and a binary velocity of  $v_S=310$\, km/s at $r=1.2$\, kpc ($\alpha=1$).
\end{itemize}

These values of $\epsilon$, which have been computed assuming a circular Keplerian orbit for simplicity, show that the acceleration of a galaxy in the potential well of its cluster usually exceeds the one of the binary due to its galactic orbit.
This situation changes if one considers stellar mass BH binaries belonging to a population segregated within 1 pc of the galactic center \cite{OLeary:2008xt}.
These sources would in fact experience a much stronger galactic acceleration due to their proximity to the MBH residing at the center of the galaxy, and would provide values of $\epsilon$ of order $10^4$ and higher.
Although for these binaries the effect of peculiar accelerations would certainly give a larger contribution to the gravitational waveform, their analysis requires a specific investigation which goes beyond the scope of the present work.
In what follows we will thus focus on stellar BH binaries with average galactic velocities and distances, resulting in the values for $\epsilon$ listed above.

Fig.~\ref{fig:LISA} shows the mismatch as a function of redshift for several values of the chirp mass at the source. 
The range in redshift for each case has been determined by the requirement that the LISA SNR is larger or comparable to 5.
Hence in the x-axes of Fig.~\ref{fig:LISA} the highest redshift corresponds to the maximal one at which a black hole binary with the mass range under analysis is visible by LISA. Note that higher masses are expected to be detected at increasingly higher redshift: we give more details about the SNR vs redshift in Appendix \ref{appendix}.
From Fig.~\ref{fig:LISA} it is clear that in the low mass region the effect of peculiar acceleration is higher, as already argued from Eqs.~\eqref{DePhiCoal} and \eqref{DePhiDet}, leading to a higher mismatch. For masses $M_c\gtrsim 100\,M_\odot$ the effect becomes negligible, even if the SNR increases.
 
For $\epsilon<50$, we observe a redshift dependence of the mismatch in
agreement with the analytical estimates of Eqs.~\eqref{DePhiCoal} and
\eqref{DePhiDet}.
The effect due to the binary peculiar acceleration dominates over the one from the background acceleration: as shown in Fig.~\ref{fig:Fz}, the former depends very mildly on redshift at low $z$. If the mass is small, the dephasing is given by Eq.~\eqref{DePhiDet}, from which it is clear that for $M_c\lesssim 50\,M_\odot$ virtually no dependence on redshift is visible at $z\lesssim 0.1$: this is what is observed in Fig.~\ref{fig:LISA} for small masses. On the other hand, when the mass is high enough ($M_c\geq 100\,M_\odot$ in Fig.~\ref{fig:LISA}), one can appreciate the redshift dependence as $(1+z)^{-10/3}$ inherited from Eq.~\eqref{DePhiCoal}. For $\epsilon=0$, we see that the mismatch increases with redshift since in this case the relevant quantity is the redshift dependence due to the effect of the background acceleration, $X(z)$.

On the other hand, for large accelerations ($\epsilon>50$) and low masses ($M_c< 50 \, M_\odot$), the points are
scattered and the redshift dependence of the mismatch displays somehow random
oscillations: this is not a physical effect, but it is due to the discreteness
of the Monte Carlo sampling and to the fact that the mismatch in this region
of the parameter space is extremely
sensitive even to tiny variation of the mass values of the template which has
$O(10^6)$ or more GW cycles in band.
A more sophisticated code is necessary to reliably identify miminum mismatch
waveforms in this region. By performing several Monte Carlo runs, we have
however verified that the values of the mismatch for $\epsilon>50$ are indeed
stable in order of magnitude also for $M_c< 50 \, M_\odot$, suggesting that
the obtained mismatches can at least be taken as an estimate.
Note that in every run, the Monte Carlo code tests for the injection point and
in most cases does not find a minimum there.

For each mass value in Fig.~\ref{fig:LISA}, $f_{\rm min}$ has been chosen within the interval given in Eq.~\eqref{fmin} in order to maximise both the mismatch and the range of redshift where the binary is visible. Fig.~\ref{fig:fmin} shows how the mismatch varies as a function of $f_{\rm min}$ for fixed $\epsilon=12$ and fixed redshift $z=0.05$, chosen for illustrative purposes. We see that the mismatch is maximised if $f_{\rm min}$ is low, since more cycles are visible before the signal exits the LISA band. However, as discussed in Appendix~\ref{appendix}, the SNR of the binaries increases with $f_{\rm min}$ (except for the largest $M_c$), see Fig.~\ref{fig:SNR}. Choosing a higher $f_{\rm min}$ allows us therefore to extend the range of redshift in which the binary in visible. With these considerations in mind, we have chosen small values of $f_{\rm min}$ for $M_c\geq 100\,M_\odot$: $0.004$ Hz and $0.006$ Hz; and a larger value, $f_{\rm min}=0.009$ Hz, for the lower masses.

In general, from Fig.~\ref{fig:LISA} we see that the typical mismatch we obtain by trying to recover signals that contain the acceleration effect with the pure TaylorF2 template (as in \eqref{Psi+} with $Y=0$) are tiny and do not exceed $10^{-3}$. This means that neglecting the acceleration effect does not generate a noticeable loss of GW detections. However, how much bias on the recovered parameters does the acceleration effect generate?

\subsection{The bias on the symmetric mass ratio and the time of coalescence}

Let us first consider the bias on the symmetric mass ratio $\eta$. On the left panel of Fig.~\ref{fig:eta} we show the fraction of events for which the injected value $\eta=0.25$ is recovered within 1\%. This is the average precision with which LISA is expected to recover the mass ratio, as shown for example in \cite{Sesana:2016ljz}. Already for $\epsilon=12$, 15\% of the events with $M_c= 30~M_\odot$ have a recovered $\eta$ wrong by more than 1\%, and the percentage raises to 40\% for $M_c= 20~M_\odot$. 

The value of the time of coalescence is also recovered with a significant bias due to the acceleration effect, as shown in Fig.~\ref{fig:eta}. The right panel represents the fraction of events with $t_c$ recovered within one minute, which is the highest uncertainty on $t_c$ for LISA reported in~\cite{Sesana:2016ljz} for the low-mass black hole binaries visible in both the LISA and LIGO/Virgo bands. Again, for the smallest value of $\epsilon=12$, 10\% of the events with $M_c=40\,M_\odot$ and more than 40\% with $M_c=20\,M_\odot$ have a recovered $t_c$ wrong by more than one minute. This will be relevant for the prediction of the time at which the binary will be visible in the LIGO/Virgo band (near merger). 

The present analysis provides us with indications that the bias on $\eta$ can be of few percent, and the one on $t_c$ can be of several days: however, a more precise investigation is in progress to determine these biases more accurately for low masses ($M_c<50 M_\odot$).

Note that in this preliminary analysis our concern is not about a very accurate
determination of the intrinsic measure uncertainty with the Monte Carlo code.
In the case of no effect, 
$Y=0$ (see Eq.~\eqref{Y}), we recover for example $\mathcal{M}_c$ at values with relative difference from the
injected ones of at most a few $\times~10^{-6}$. This remains the case when $Y\neq 0$.
Comparing with the back of the envelope calculation performed in section \ref{estimate} (c.f Eq.~\eqref{dePHIetaconst}) 
one can conclude that no relevant bias is induced on $\mathcal{M}_c$ by the total acceleration effect. 
However, we can deduce from Eq.~\eqref{dePHIMcconst} and e.g.~\cite{Sesana:2016ljz} that the uncertainty
due to detector noise on
$\eta$ and $t_c$ is smaller than the bias we find here for a large number of
events, especially at low masses. Hence we are confident that our result holds in the
presence of intrinsic measure uncertainties.

\section{Conclusions}
\label{se:conclusion}

We have analysed the effect of redshift perturbations on the GW form. GWs emitted by inspiral binaries propagate through the inhomogeneous universe before reaching the detector. These inhomogeneities influence the observed frequency of the GW: in addition to the background redshift due to the expansion of the universe, the inhomogeneous distribution of matter generates a Doppler shift (due to the peculiar velocity of the binary with respect to the observer), a gravitational shift and an integrated Sachs-Wolfe effect. We have found that if the redshift perturbations are constant during the time of observation of the GW, the waveform does not change: the redshift perturbations can simply be reabsorbed into the redshifted chirp mass $\mathcal{M}_c(z)$. On the other hand if the redshift perturbations evolve during the time of observation, they generate a contribution to the waveform with frequency dependence $f^{-13/3}$ (formally a -4PN term).

We have compared the amplitude of this novel effect with the correction
generated by the background acceleration of the universe, derived previously in~\cite{Seto:2001qf, Nishizawa}. The dominant correction from redshift perturbations comes from the peculiar acceleration of the binary (i.e.~the variation of the peculiar velocity during the time of observation). We found that this contribution from the peculiar acceleration dominates, for realistic situations, over the background expansion one over a large range of redshift. As we do not know in practice what is the amplitude of the peculiar acceleration for individual binaries, this strongly challenges the possibility of using the background effect to determine the redshift of the binary.

We have then performed a preliminary analysis of the impact of the binary peculiar acceleration on the recovery of the binary parameters for the LISA detector,
using inspiral-only, spin-less waveforms as a test case.
The effect is most relevant for low-mass binaries with source chirp mass
$M_c<100\,M_\odot$ at low redshift and entering the detector at frequency around
few mHz. For these kind of sources, the total phase shift due to the acceleration effect is 
comparable to the one induced by high PN terms: 2PN order at best.

We have found that using a template without the acceleration effect to analyse GW signals does not cause significant loss of detections since the mismatch is at most of $10^{-3}$. However, our results show that the recovered parameters are biased by the acceleration effect. Although this does not happen for the redshifted chirp mass $\mathcal{M}_c$, we have found that for a large fraction of events at small masses, $\eta$ is not recovered within 1\% and the time of coalescence $t_c$ is wrong by more than 1 minute (we compare with the estimated errors with which LISA is expected to measure these quantities, according to \cite{Sesana:2016ljz}). Our analysis provides us with indications that the bias on $\eta$ can be of few percent, and the one on $t_c$ can be of several days: however, a more precise investigation is in progress to determine these biases more accurately at low masses
(source chirp masses less than $50 M_\odot$).
The estimate of $t_c$ is of particular significance for the binaries that can be detected first by LISA and then by ground based interferometers for which one needs a precise determination of the coalescing time.

In order to remove this bias one should add to the GW templates a dependence on
the acceleration effect and introduce a new search parameter $\epsilon$ 
(see Eq.~\eqref{epsilon}).
This procedure might reduce the precision with which the parameters of the
binary are recovered but it could possibly allow a measurement of the peculiar
acceleration of the binary, which may convey valuable information about the
environment in which the binary is living.

\section*{Acknowledgements}
We would like to thank Enrico Barausse, Daniel Holz, Sean McWilliams, Bangalore Sathyaprakash and Alberto Sesana for very useful discussions, and Zoltan
Haiman for pointing out to us the implications of Ref.~\cite{OLeary:2008xt}. CB acknowledges support by the Swiss National Science Foundation. RS acknowledges the support of Fundac\~ao de Amparo \`a Pesquisa do estado de S\~ao Paulo through grants 2012/14132-3 and 2013/04538-5 and CERN for hospitality and support during the early stage of this work.
This research was supported by resources supplied by the Center for Scientific
Computing (NCC/GridUNESP) of the S\~ao Paulo State University (UNESP).
CC and NT thank the {\it Institut d'Astrophysique de Paris}, the institute {\it AstroParticule et Cosmologie} at {\it Universit\'e Paris-Diderot}, and {\it CERN} for hospitality during the completion of this work. 

\appendix
\section{Signal to noise ratio}
\label{appendix}

\begin{figure*}[t]
\centering
\includegraphics[width=18cm]{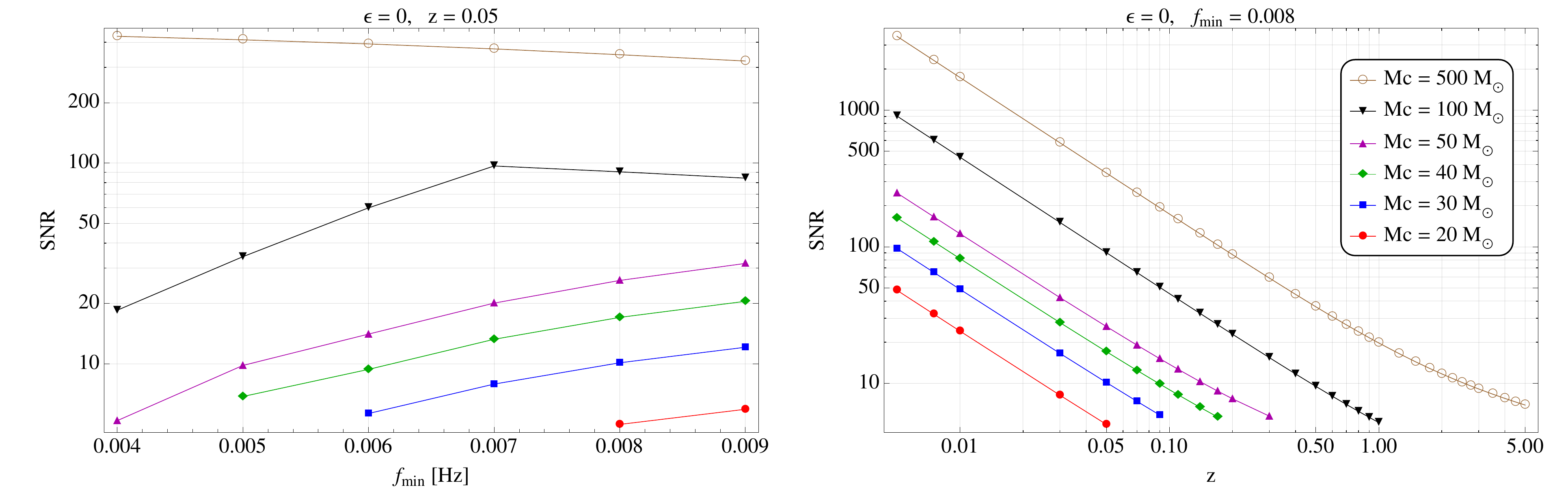}
\caption{SNR as a function of redshift and $f_{\rm min}$ for different chirp masses. To compute the SNR we set $\epsilon=0$, i.e.~the injected signal coincides with the template.}
\label{fig:SNR} 
\end{figure*}

In this appendix we discuss the LISA SNR for the events considered in the main text.
The results are summarised in Fig.~\ref{fig:SNR}: the left panel shows the SNR as a function of $f_{\rm min}$ for events at fixed $z=0.05$ (chosen to be able to display enough points in redshift for low masses), while the right panel shows the SNR as a function of redshift for events with $f_{\rm min}=0.008$ Hz.
To compute the SNR we set $\epsilon=0$, i.e.~the injected signal is assumed to coincide with the template:
${\rm SNR}=\sqrt{\langle h_2|h_2\rangle}$. 
As shown in Fig.~\ref{fig:SNR}, within the mass range used in our analysis the higher the masses the higher the SNR.
The left panel shows that the SNR always grows with $f_{\rm min}$ for low mass binaries ($M_c \leq 50 M_\odot$), while it reaches a maximum within the displayed frequency range for $M_c =100 M_\odot$, and always decreases for $M_c =500 M_\odot$. This behaviour is due to the fact that, keeping the masses and distance fixed, binaries with higher $f_{\rm min}$, i.e.~which appear at higher frequencies when the detector is turned on, are in a more relativistic phase of their evolution,
with a smaller separation distance and consequently a stronger GW emission.
However, binaries with higher masses entering at the same $f_{\rm min}$ evolve faster over the frequency range and exit the LISA band before the end of the five years observation time: the SNR is therefore reduced.
This happens for chirp masses roughly above $100\, M_\odot$, as shown in Fig.~\ref{fig:SNR}.
The behaviour of the SNR is simpler in terms of the redshift of the sources, as displayed in the right panel of Fig.~\ref{fig:SNR}.
As expected, the higher the redshift the smaller the SNR, since the GW signal
is inversely proportional to the distance.

\bibliographystyle{h-physrev4}
\bibliography{bibfile_gw_pert}

\end{document}